\documentclass[twocolumn,aps,10pt,showpacs,showkeys,]{revtex4}
\usepackage[dvips]{graphicx}

\begin{document}
\title{Crystal field and spin-fluctuations \\ in heavy fermion metal CeAl$_{3}$, YbRh$_{2}$Si$_{2}$ and
UPd$_{2}$Al$_{3}$$^\spadesuit$}
\author{R. J. Radwanski}
\affiliation{Center of Solid State Physics, S$^{nt}$Filip 5, 31-150 Krakow, Poland,\\
Institute of Physics, Pedagogical University, 30-084 Krakow,
Poland}
\author{Z. Ropka}
\affiliation{Center of Solid State Physics, S$^{nt}$Filip 5,
31-150 Krakow, Poland} \homepage{http://www.css-physics.edu.pl}
\email{sfradwan@cyf-kr.edu.pl} This paper has been presented at
SCES-05

\begin{abstract}
We have shown the existence of localized crystal-field states in
conducting magnetic materials coexisting with conduction
electrons in CeAl$_3$, UPd$_2$Al$_3$ and YbRh$_2$Si$_2$. We point
out that the magnetism is strongly correlated with the very
detailed local atomic-scale crystal structure.

\end{abstract}
\maketitle

 30 years after the discovery of an anomalous behavior of
CeAl$_{3}$ marked by an enormously large specific heat at low
temperatures, with the Sommerfeld coefficient exceeding
1600~mJ/K$^2$mol \cite{1} the origin of the heavy-fermion (h-f)
behavior is still unclear. CeAl$_3$, UPd$_2$Al$_3$ and
YbRh$_2$Si$_2$ are all metallic compounds. The aim of this paper
is to point out the importance of local crystal-field (CEF)
interactions in these metallic compounds owing to the fact that
still the concept and the use of the crystal-field approach is
questioned, if applied to metallic compounds. There are opinions
that obtained agreements with experimental data within the
CEF-based approaches are accidental. For us the many-electron CEF
approach is obvious and well founded in solid-state physics. It,
keeping the atomic-like integrity of the $4f/5f$ shell, captures
the essential physics of transition-metal compounds. The CEF
results from the inhomogeneous charge distribution in a solid so
it is everywhere. Another problem is its quantification. In a
perfect crystal due to the perfect translational symmetry and a
high point symmetry the CEF potential can be quantified by CEF
coefficients that reflect multipolar charge moments of the
lattice. A transition-metal cation with the incomplete $4f/5f$
shell serves as a probe of this CEF multipolar potential showing
the discrete splitting of the localized $4f/5f$ states.
\begin{figure*}[t]
\begin{center}
\includegraphics[angle=270,width = 10.5 cm]{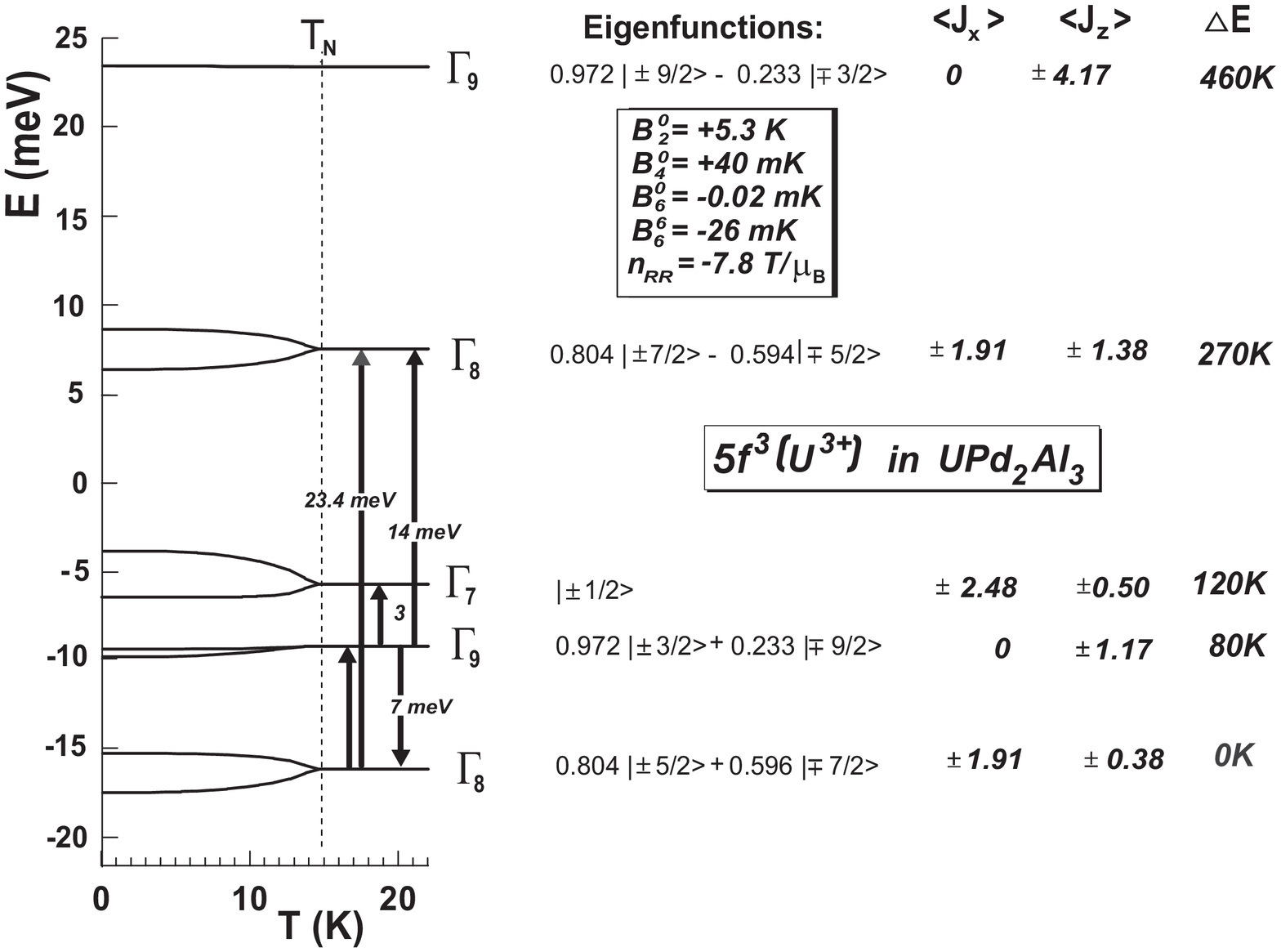}
\end{center} \vspace {-0.3 cm}
\caption{Calculated energy level scheme of the $5f^3$
configuration in UPd$_2$Al$_3$. Arrows indicate transitions which
we have attributed to excitations observed by
inelastic-neutron-scattering experiments of Krimmel et al.
\cite{2}.}
\end{figure*}

All mentioned compounds have an incomplete $f$ shell with an odd
number of electrons ($f^1$ in CeAl$_3$, $f^3$ in UPd$_2$Al$_3$,
$f^{13}$ in YbRh$_2$Si$_2$). As the effect of multipolar charge
interactions (=CEF interactions) in a solid , in all cases the
ground state is the Kramers doublet. Such the Kramers doublet is
unstable with respect to spin-dependent interactions that have to
remove the double degeneracy before the system reaches T =0~K.
The removal of the Kramers degeneracy is equivalent to the
breaking of the time-reversal symmetry and to the formation of a
magnetic state, less or more complex. Due to the presence of the
substantial orbital moment ($f$ electrons) all of these compounds
can serve as example of anisotropic spin lattices with quite low
ordering temperatures.

UPd$_2$Al$_3$ exhibits supercoductivity below 2 K coexisting with
antiferromagnetism that appears below 14 K. We have interpreted
excitations observed by Krimmel {\it et al.} \cite{2} as related
to the energy level scheme: 0, 7 meV (81 K), 10 meV (116 K) and
23.4 meV (271 K) of the $f^{3}$ (U$^{3+}$) configuration
\cite{3,4}. The fine electronic structure of the $f^{3}$
consisting of five Kramers doublets originating from the lowest
multiplet $^{4}$I$_{9/2}$ is shown in Fig. 1. The Kramers
doublets are split in the antiferromagnetic state, i.e. below
T$_N$ of 14 K. We are convinced that a low-energy excitation of
1.7 meV at T=0 K observed by Sato {\it et al.} \cite{5} and
ascribed to a magnetic exciton, is associated to this energy
splitting of the Kramers-doublet ground state in the
antiferromagnetic state. Its observation confirms our
interpretation with the $5f^3$ configuration. The spin-dependent
interactions produce the magnetic order below T$_N$ \cite{3,4}
what is seen in Fig. 1 as the appearance of the splitting of the
Kramers doublets and in experiment as the $\lambda$-peak in the
heat capacity at T$_{N}$. This electronic structure accounts for
the overall temperature dependence of the heat capacity, the
substantial uranium magnetic moment and its direction \cite{4}. We
would like to point out that the derived ground state is formed
by higher-order CEF interactions.

YbRh$_2$Si$_2$ being by years one of a hallmark heavy-fermion
compound with a Kondo temperature T$_K$ of 25-30~K becomes
recently of great scientific importance after the discovery in
Prof. F. Steglich group \cite{6} of the well-defined Electron Spin
Resonance (ESR) signal at temperature of 1.5~K, i.e. much below
T$_K$. Its observation violates the Kondo model according to
which there should not be discrete states below T$_K$. The derived
very anisotropic $g$ tensor: $g_{\bot }$~=~3.561 and $g_{\Vert
}$~=0.17 at 5~K we have described as related to the CEF
interactions. In Ref. \cite{7} we have derived two sets of CEF
parameters with Kramers doublet ground-state eigenfunctions,
$\Gamma _{6}$ or $\Gamma _{7}$:

$\Gamma _{6}$ = 0.944 $|\pm 1/2>$ + 0.322 $|\mp 7/2>$

$\Gamma _{7}$ = 0.803 $|\pm 3/2>$ + 0.595 $|\mp 5/2>$

These ground states reproduce the observed $g$ tensor perfectly
well, particularly well if we allow, apart of the tetragonal local
symmetry, a small admixture, below 0.5$\%$, of functions due to
the orthorhombic distortion B$_2^2$. A Mossbauer experiment,
revealing temperature dependence of the local quadrupolar
interactions, that can distinguish between these two ground
states, is still awaited. The crystal-field approach yields large
anisotropy of magnetic properties - in Fig. 2 we present the
temperature dependence of the paramagnetic susceptibility
$\chi(T)$ calculated for different crystallographic directions for
the $\Gamma _{7}$ ground state. A huge anisotropy of $\chi(T)$ is
consistent with the anisotropy of the $g$ tensor and experimental
data shown in Fig. 1 of Ref. \cite{8}. The larger susceptibility
in the tetragonal plane than along the tetragonal axis is
consistent, within the single-ion CEF model, with the moment
direction $\Vert $c observed in antiferromagnet CeRh$_2$Si$_2$
below T$_{N}$=35~K (the second-order Stevens factors are opposite
for $f^{1}$ and $f^{13}$).
\begin{figure}[ht]
\begin{center}
\includegraphics[width = 9.4 cm]{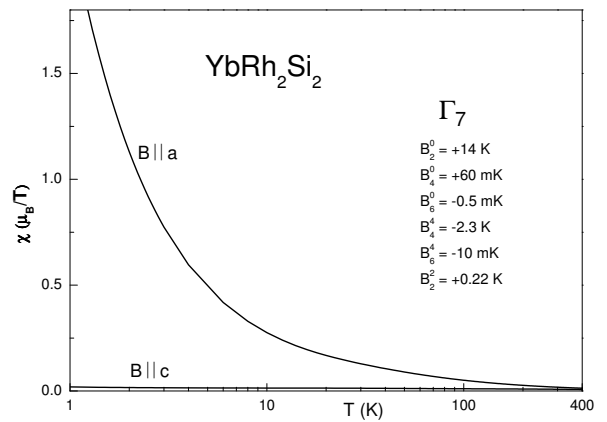}
\end{center} \vspace {-1.5cm}
\caption{Calculated temperature dependence of the paramagnetic
susceptibility of YbRh$_{2}$Si$_{2}$ for the $\Gamma _{7}$ ground
state. } \label{radwanski-633-f2. }
\end{figure}

A magnetic state of CeAl$_3$ has been found only in single
crystalline sample in 1993 \cite{9} almost 20 years after the
first report in 1975 on exotic low-temperature properties, later
called as heavy-fermion phenomena. Lapertot {\it et al} \cite{9}
found that low-temperature magnetic and heat properties are very
sensitive to the exact stoichiometry and the sample quality and
found well pronounced peak in c(T) marking magnetic order at
T$_N$ of 2.16~K. An inelastic neutron scattering experiment of
Alekseev {\it et al} \cite{10} reveals a CEF-like excitation at 7
and 15 meV. These excitations can be described by a set of CEF
parameters of the trigonal symmetry: B$_2^0$= -3.2~K,
B$_4^0$=+0.51~K and B$_4^3$=+1.5~K yielding the $ |\pm3/2\rangle$
ground Kramers doublet.

In conclusion, we have shown the existence of localized
crystal-field states in conducting magnetic materials coexisting
with conduction electrons. We point out that magnetism is
strongly correlated with the very detailed local atomic-scale
crystal structure.\\

$^\spadesuit$ Dedicated to the Pope John Paul II, a man of
freedom and truth in life and in Science.

\end{document}